# Situationally-Induced Impairments and Disabilities Research

**Zhanna Sarsenbayeva**
School of Computing and Information Systems
The University of Melbourne, Australia
zhanna.sarsenbayeva@unimelb.edu.au

**Vassilis Kostakos**
School of Computing and Information Systems
The University of Melbourne, Australia
vassilis.kostakos@unimelb.edu.au

**Jorge Goncalves**
School of Computing and Information Systems
The University of Melbourne, Australia
jorge.goncalves@unimelb.edu.au

## ABSTRACT

Research has shown that various environmental factors impact smartphone interaction and lead to Situationally-Induced Impairments and Disabilities. In this work we discuss the importance of thoroughly understanding the effects of these situational impairments on smartphone interaction. We argue that systematic investigation of the effects of different situational impairments is quintessential for conducting successful research in the field of SIIDs that might lead to building appropriate sensing, modelling, and adapting techniques. We also provide insights for future work identifying potential directions to conduct research in SIIDs.

## INTRODUCTION

As smartphones have become an indispensable part of our everyday life, it is increasingly common to use them under various contextual factors that impair mobile interaction. Impaired abilities to interact with the device due to contextual factors are known as Situationally-Induced Impairments and Disabilities (SIIDs) or situational impairments [9].

A decade ago Wobbrock [11] envisioned that a better understanding of situational impairments could also lead to a better understanding of permanent impairments as contextual factors could impair mobile device user in a similar way as cognitive or physical impairments affect users with disabilities [1]. This vision has since been showcased in different studies. For example, Yesilada *et al.* [12] demonstrated that a situationally-impaired smartphone user performed a similar number of errors as a physically-impaired user on a desktop. Therefore, designing solutions that address situational impairments can also potentially benefit permanently impaired users.

Furthermore, research has also shown that SIIDs aggravate the experience of permanently-impaired users during mobile interaction [3]. For example, a visually impaired mobile device user, walking in an unfamiliar environment, undergoes an endeavour of dividing attention between swiping the cane, navigating in the environment, and interacting with the smartphone. Therefore, it is important to consider SIIDs when designing accessible technology.

Research conducted in this area falls within four main categories: *Understanding*, *Sensing*, *Modelling*, and *Adapting* [10]. In our work we focus mostly on *Understanding*, as it is necessary to accumulate and build knowledge regarding the impact of different SIIDs on mobile interaction before developing the necessary sensing technologies, or adapting interfaces to accommodate the sensed impairments.

A fair amount of research has been conducted to study and understand the impact of SIIDs on mobile interaction; however, the majority of the studies follow an ad-hoc approach and lack systematic investigation. This results in the inability of the research to draw a comparative judgement of the magnitude of the effect of different SIIDs on mobile interaction.

In this position paper we elaborate on why it is important to conduct systematic research on SIIDs, and discuss how further progress in SIID research depends on a thorough *Understanding* of the effects of various situational impairments on mobile interaction.

## UNDERSTANDING SIIDS IMPACT ON MOBILE INTERACTION

We consider *Understanding* to be the quintessential step that empowers conducting successful research in SIIDs. It is important to understand the exact effects contextual factors have on mobile interaction. Once the effect is known, it introduces possible sensing, modelling, and adapting mechanisms to address the particular situational impairment without wasting resources.

For example, our early work [2, 4] established the negative effect of cold ambience on mobile interaction. We show that in a cold setting participants take longer time to access targets, as well as they become less accurate in target acquisition tasks. Hence, as the effect of cold ambience was significant, there was a need to develop an effective mechanism to sense cold-induced situational impairments, as such a mechanism did not exist beforehand. Hence, in our following work [8] we introduced a mechanism to sense cold-induced situational impairments using mobile phone's battery temperature.

Not only is it important to establish the effects of various SIIDs on mobile interaction, but it is crucial to be able to compare these effects between each other [6]. While other studies have attempted to understand the impact of different SIIDs on mobile interaction in an ad-hoc fashion, we have tackled this challenge in a systematic way. Namely, we have developed an experimental protocol to be reused across multiple studies. In particular, we have been re-using same smartphone tasks to quantify the effect of different situational impairments on mobile interaction: target acquisition, visual search, and text entry. We argue that these three tasks are the most fundamental in mobile interaction. We also restrict our participants posture for holding smartphone device to perform two-handed interaction: holding a smartphone in a non-dominant hand, while interacting with the smartphone with an index finger of a dominant hand. Furthermore, in each of the studies we set a respective baseline condition to observe mobile interaction without the effect of situational impairment. In addition, we provide our participants with extensive training of the tasks before the experiment starts to avoid any possible sequence effects.

All of the aforementioned regulations yield fair conditions to apply further equitable comparison of the effects of various situational impairments in terms of how disruptive they are during mobile interaction. This then enables us to contrast the effects of different situational impairments in terms of percentile growth/drop as compared to the respective baseline for the different types of tasks.

This direct comparison of the effects can facilitate building appropriate sensing, modelling, and adapting mechanisms that would accommodate the most prominent situational impairment while at the same time potentially addressing other accompanying SIIDs. For example, as a result of our systematic approach [5], we can see that the effect of stress is less prominent than the effect of ambient noise (meaningful speech) on the text entry tasks, while being more prominent in other types of tasks. This is a good showcase of meaningful speech impairing participants' ability to type, as they had to divide their attention between listening to the speech and thinking what to type. In case of both of these situational impairments being sensed, the device should choose to address noise-induced situational impairment in case the user is performing text entry tasks, while accommodating any accompanying stress-induced situational impairment in other tasks.

## CONCLUSION AND FUTURE WORK

In this paper we discuss the importance of conducting empirical and systematic research to understand the effects of situational impairments on mobile interaction. We show that understanding the effects of SIIDs is the pillar of SIID research, as it enables further construction of sensing mechanisms, creation of techniques to model SIIDs and user behaviour under SIIDs, as well as interface adaptation to accommodate for SIIDs.

We envision that future research should focus on investigating the effects of combined situational impairments. For example, how would the combination of cold ambience together with outdoor urban noise affect mobile interaction. We show in our previous work that participants take significantly longer time to tap targets when exposed to cold environment [4]; however, we also show when exposed to urban outdoor noise, participants' target acquisition time significantly decreases [7]. In the case of combining these two situational factors, it is unknown if their effects would cancel each other or if one of the factors would outperform the other one. However, we emphasise that the effects of combined situational factors should be studied systematically.

Future work should also investigate more complex smartphone tasks to determine the effects of contextual factors on physical and cognitive performance during mobile interaction. In our research we limited our studies to the three common smartphone tasks: target acquisition, visual search, and text entry. We used same tasks throughout our studies to maintain systematic approach to understand the effects of situational factors. However, the effect of some situational factors was not significantly profound on particular tasks. For example, in a study where we quantify the effects of ambient noise on mobile interaction, we show that meaningful speech (speech that the user can understand) had a significant negative effect on the text entry task; however, the effect of stress was not significant. We argue that using more complex tasks would better describe the effect of the situational factor, particularly when dealing with situational impairment that have a stronger effect on cognition.

Future research could also explore building detection mechanisms that sense the presence of a combination of situational impairments (*e.g.,* when the device is being used in a noisy street in winter). Finally, the adapting techniques should be sufficiently intelligent to accommodate the most severe situational impairments depending on the type of task the user is performing.